\documentstyle[mlpp,epsf,10pt]{article}
\begin{document}
\begin{article}
\twocolumn
\begin{flushleft}
\bf The Galactic Environments of Cool Stars -- part I: Modeling
Interstellar Dust around the Sun and Nearby Cool Stars
\end{flushleft}
\begin{center}
\small M. Landgraf$^1$ and P.C. Frisch$^2$
\end{center}
\begin{flushleft}
\tiny
$^1$ NASA/JSC, Houston, TX 77058\\
$^2$ University of Chicago, Chicago, IL 60637\\
\end{flushleft}
\begin{abstract}
We present a model of the interaction of interstellar dust grains with
a stellar environment, that predicts the distribution of interstellar
dust grains in the size range between $0.1\;{\rm \mu m}$ and $1\;{\rm
\mu m}$ around a star for the whole stellar cycle. Comparisons of the
model results with in-situ dust measurements by the Ulysses spacecraft
in the Solar System validate the model. We show that in the case of
the Sun, interstellar dust grains can produce large regions of
infrared emission that can be confused with a circumsolar dust disk
when observed from afar. Our model can determine the shape of
interstellar dust concentrations close to nearby stars, if we have
information on the relative velocty of the stars with respect to the
surrounding interstellar medium, and the properties of the stellar
wind.
\end{abstract}

\section{Introduction}
the stellar wind deflects interstellar dust grains from their initial
direction of motion. The actual direction in which the grains are
deflected depends on the polarity of the stellar magnetic field and
thus on the phase of the stellar cycle. Since the magnetic field
produced by the star is usually stronger than the surrounding
interstellar magnetic field, the filtration at the heliopause affects
smaller grains than the filtration inside the heliosphere. Both
filtration mechanisms disturb the flow of interstellar grains around
the star. Therefore, interstellar dust density fluctuations can be
expected on scales of the stellar astrosphere. This has to be taken
into account when infrared observations of the stellar vicinity are
analyzed, because spatial disturbances of the interstellar dust
distribution can mimic the appearance of circumstellar dust disks.

Another aspect of the interaction of interstellar dust with the
stellar environment is the erosion of circumstellar dust disk by
collisions of disk particles with interstellar grains. It has been
shown [Liou and Zook, 1999] that the appearance of a dust disk can be
a strong indicator of planets that orbit the star. The indicative
features of such a dust disk develop on time scales in the order of
$10^6$ to $10^7$ years. Depending on the concentration and relative
velocity of interstellar dust near the star, these features are eroded
before they become strong enough to be detected.


\section{Model description}
We simulate the gravity, radiation pressure, and Lorentz-force that
interstellar grains experience as they traverse the heliosphere and
the Solar System (for a full description see Landgraf [1999]). The
relative strength of these forces depends on the grains' size. Grains
with sizes below $0.1\;{\rm \mu m}$ experience dominantly the
Lorentz-force that is created by the solar wind magnetic field
sweeping by the grains [Morfill and Gr\"un, 1979]. Since the solar wind
magnetic field changes its polarity with the solar cycle, the effect
on the spatial grain distribution also depends on the phase of the
cycle. Gr\"un et al. [1994] and Gustafson and Lederer [1996] found that
during the first half (1991 to 2002) of the cycle, grains are
deflected away from the solar equatorial plane, and towards the solar
equatorial plane during the second half (2002 to 2013). If the grain
size is comparable to the maximum wavelength of the solar spectrum,
radiation pressure becomes effective and can even exceed gravity
[Gustafson, 1994]. The motion of grains with sizes above $1\;{\rm \mu
m}$ is dominated by gravity and they approximately move on hyperbolic
Kepler orbits.

\section{Simulation results}
\begin{figure}[ht]
\epsfxsize=\hsize
\epsfbox{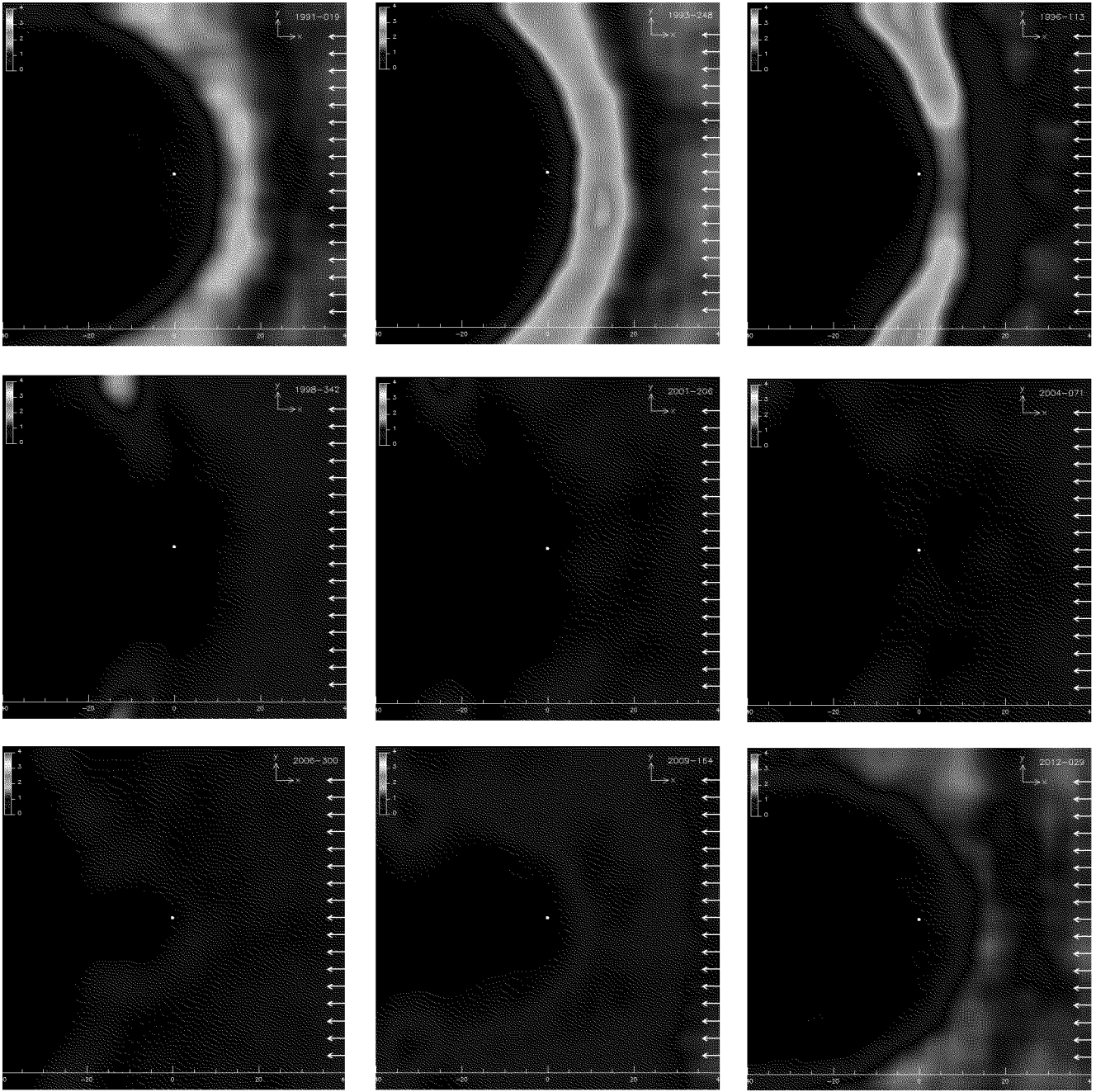}
\caption{\label{fig_panelxy}\it Spatial distribution of interstellar
grains with radii of $0.1\;{\rm \mu m}$ in a $80\;{\rm AU} \times
80\;{\rm AU}$ plane that contains the upstream direction vector and is
close to the solar equatorial plane. The distribution changes with
the solar cycle and is therefore shown as a sequence (from upper left
to lower right) of panels that covers the whole cycle from 1991 to
2013.}
\end{figure}

In figure \ref{fig_panelxy} we show the spatial
distribution of spherical interstellar dust grains with radii of 0.1
microns around the Sun during the whole solar cycle from 1991 to
2013. For this simulation we have assumed an initial relative velocity
of the grains of 26 km/s with respect to the Sun and a electrostatic
grain charge surface-potential of +5V [Mukai, 1981]. The figure
shows how the spatial distribution of interstellar dust grains around
the Sun changes with the 22-year solar cycle. The individual paneles
in figure \ref{fig_panelxy} show a rectangular
area of 80AU x 80AU around the Sun. Figure \ref{fig_panelxy} shows the
distribution in the plane that contains the initial dust velocity
vector and is perpendicular to the plane that contains the solar
rotation axis and the dust velocity vector. In the case of the Sun,
the plane shown in figure \ref{fig_panelxy} is also 
close to the solar equatorial plane. During all phases of the solar
cycle the solar wind magnetic field repels interstellar dust grains
from the vicinity of the Sun. The average magnetic field strength
experienced by a dust grain is higher when the field is highly ordered
during a solar minimum. The first panels show the distribution of
interstellar dust grains after they have been concentrated to low
heliographic latitudes by a field configuration that deflects
particles in the northern hemisphere to the south and particles in the
southern hemisphere to the north during the 1985 solar minimum. The
arc of enhanced particle density downstream of the Sun is generated by
the repelling effect of the solar wind magnetic field. Since the
average field strength is low during the solar maximum in 1991, the
particles propagate freely until they are deflected to high
heliographic latitudes during the 1997 solar minimum. After the
projected solar maximum in 2002 particles are deflected to lower
heliographic latitudes again, and in 2012 the same distribution as at
the beginning of the cycle in 1991 is achieved.

\section{Comparison with Spacecraft Measurements}
Ulysses has measured a decrease in the flux of interstellar grains
after mid-1996 by about a factor of 3. Our model reproduces this
decrease, which is due to the deflecting phase of the solar wind
magnetic field that starts in 1991 and becomes most effective in 1995,
according to the model interpretation. Consequently the spatial
concentration of interstellar grains decreases, especially around low
heliographic latitudes. Since the grain's charge-to-mass ratio, that
controls the coupling of the grain to the magnetic field, is inversely
proportional to the grain radius, the smallest grains are affected
most by the deflecting field. Larger grains are depleted more slowly,
due to their inertia. The best fit to the data is achieved for grain
sizes between $0.2\;{\rm \mu m}$ and $0.3\;{\rm \mu m}$, which is also
the size range of the most abundant impacts onto the Ulysses dust
instrument before mid-1996 [Landgraf et al., 1999]. We conclude that
the in-situ measurements support the model prediction that the
distribution of small interstellar dust grains around the Sun is
time-dependent and changes with the solar cycle.

\section{Prediction of Interstellar Dust around Nearby Stars}
Can we predict the spatial distribution of interstellar dust around
nearby stars?
To model the motion of interstellar dust grains in the vicinity of a
nearby star we need information on the relative velocity of the star
with respect to the surrounding medium, the magnetic cycle, and the
velocity and thermodynamic state of the stellar wind. To assess dust
filtration at the astrospheric boundary region, information on the
magnetic and thermal pressure in the surrounding medium is
needed.

The Hipparcos catalog contains 1550 stars within $25\;{\rm pc}$, 508 of
which are G-type stars. The astrometric data can be used to determine
the star's velocity. The velocity of the surrounding medium can be
determined from absorption line measurements of the nearby
interstellar gas [Lallement and Bertin, 1992; Frisch 1996]. The state
and periodicity of the stellar wind can be derived from measurements
of the stellar mass-loss and activity. Our model is restricted to
stars that have a magnetic field configuration similar to the Sun. The
motion of very small grains inside an astrosphere can not easily be
modeled if the grain's gyration frequency is in the order of or less
than the rotation frequency of the star. In the case of the Sun this
is not a strong restriction, because grains with sizes below
$0.05\;{\rm \mu m}$ are filtered at the heliopause region and thus
never reach the inner part of the heliosphere.

The result of our model calculations is a three-dimensional
distribution of grains around a given star. From this information
the grain flux near a circumstellar dust disk as well as the projected
distribution of infrared emission can be derived.

\end{article}

\clearpage

\end{document}